\documentclass[a4paper, twocolumn, 11pt, unpublished]{quantumarticle}
\pdfoutput=1
\usepackage[utf8]{inputenc}
\usepackage[T1]{fontenc}
\usepackage{amsmath}
\usepackage{hyperref}
\usepackage{tikz}
\usepackage{lipsum}
\usepackage{mathrsfs}
\usepackage{bbold}
\usepackage[numbers,sort&compress]{natbib} 

\begin{document}

\title{Dual channel multi-product formulas}

\author{Seung Park}
 \affiliation{Institute for Convergence Research and Education in Advanced Technology, Yonsei University, Seoul 03722,
Republic of Korea}
\affiliation{Department of Quantum Information, Graduate School, Yonsei University, Seoul 03722, Republic of Korea}
\author{Sangjin Lee}
\email{sangjin5190@gmail.com}
\affiliation{Quantum Universe Center, Korea Institute for Advanced Study, Seoul 02455, South Korea}
\author{Kyunghyun Baek}
\email{k.baek@yonsei.ac.kr}
\affiliation{Institute for Convergence Research and Education in Advanced Technology, Yonsei University, Seoul 03722,
Republic of Korea}
\affiliation{Department of Quantum Information, Graduate School, Yonsei University, Seoul 03722, Republic of Korea}
\maketitle

 \begin{abstract}
    Product-formula (PF) based quantum simulation is a promising approach for simulating quantum systems on near-term quantum computers. Achieving a desired simulation precision typically requires a polynomially increasing number of Trotter steps, which remains challenging due to the limited performance of current quantum hardware. To alleviate this issue, post-processing techniques such as the multi-product formula (MPF) have been introduced to suppress algorithmic errors within restricted hardware resources. 

    In this work, we propose a dual-channel multi-product formula that achieves a two-fold improvement in Trotter error scaling. As a result, our method enables the target simulation precision to be reached with approximately half the circuit depth compared to conventional MPF schemes. Importantly, the reduced circuit depth directly translates into lower physical error mitigation overhead when implemented on real quantum hardware. We demonstrate that, for a fixed CNOT count as a measure of quantum circuit, our proposal yields significantly smaller algorithmic errors, while the sampling error remains essentially unchanged.
\end{abstract}

\section{Introduction}
A quantum computer is an efficient platform for exploring a tremendous Hilbert space by exploiting quantum mechanics represented by superpositions and entanglements. One of the prominent applications enabled by these properties is a quantum simulation~\cite{lloyd1996universal,qsim_general3,qsim_general,qsim_appl2}, which addresses classically intractable dynamics of the quantum many-body systems in an exponentially large Hilbert space. 

The development of efficient quantum algorithms for simulating quantum many-body systems is a central topic in the field of quantum computation~\cite{Georgescu2014_qs_review, Miessen2023_qs_review}. 
To this end, a variety of state-of-art quantum simulation algorithms have been proposed, such as product-formula (PF) approaches~\cite{lloyd1996universal, berry2007efficient, childs2019nearly, Childs2021}, linear combination of unitaries methods~\cite{qsim_algo_LCU,qsim_algo_LCU2}, quantum signal processing~\cite{qsim_algo_QSP}, quantum walk-based algorithms~\cite{qsim_algo_quantumwalk,qsim_algo_quantumwalk2,qsim_algo_quantumwalk3,qsim_algo_qwalk}, and quantum singular value transformation~\cite{Martyn2021_gradn_unification}. 
Each of these approaches exhibits distinct characteristics, and thus the choice of algorithms depends on problem-specific features, such as the structure and locality of the Hamiltonian, as well as the algorithmic overhead and hardware constraints.

PF approaches \cite{lloyd1996universal, berry2007efficient, childs2019nearly, Childs2021} are particulalry promising for near-term quantum devices, as they avoid the need for controlled time-evolution or block-encoding constructions. 
For this reason, they are expected to be comparatively robust against physical errors and more readily implementable under realistic hardware constraints, such as limited coherence time and restricted qubit connectivity.
To achieve higher precision quantum simulation within the PF framework, a recursive method that generates a quantum circuit with arbitrarily high precision have been proposed  ~\cite{PF_ruth,PF_ruth2,Suzuki2,suzuki1991general,PF_nonsymmetric}. However, due to their recursive structure, these methods lead to quantum circuits whose depth grows exponentially with the desired precision.

To alleviate the issue of long quantum gate sequences, multi-product formula (MPF) methods, 
which rely on classical post-processing of measurement outcomes from multiple quantum circuits, 
have been proposed at the cost of additional resources~\cite{TEP,MPF_ref_2,MPF_ref_3,MPF_ref_4,low2019well,MPF_extrapol}. 
The primary bottleneck in implementing these algorithms arises from physical errors accumulated during circuit execution. Consequently, a key advantage of an effective algorithmic error mitigation strategy lies in achieving the desired accuracy with reduced circuit depth.

In this work, we propose an improved MPF scheme that combines the dual-channel of MPF circuits to enhance the simulation accuracy while reducing the circuit depth. 
With the proposed method, the algorithmic mitigation performance is nearly double compared to the conventional MPF approach~\cite{vazquez2023well}, for a given circuit depth and using the same number of measurements. 

The organization of this work is as follows. In the next section, we briefly review the conventional MPF method by introducing the setups of quantum simulation we are considering. In Section~\ref{sec:dcmpf}, we propose our protocols, which we dub as dual-channel MPF: it leverages dual types of PF.  We also provide a sketch of the proof of their algorithmic performances, and the details of the proof can be found in Appendix~\ref{pfDCMPF}.
In Section~\ref{sec:numerics}, we present numerical simulations demonstrating the ideal performance of our proposal as well as its performance on noisy near-term quantum simulators. Lastly, the conclusion of this work is presented in Section~\ref{sec:conclusion}.

\section{Brief review of MPF}
The product formula (PF) \cite{lloyd1996universal, suzuki1991general, berry2007efficient} is a well-known technique in quantum simulation to approximate the dynamics of a quantum system, governed by the time evolution operator {$U(t)= e^{-iHt}$} for a given Hamiltonian $H$. By decomposing the Hamiltonian into $k$-local Hamiltonians $H_j$, such as $H=\sum_{j=1}^N H_j$, a $\alpha$-th order PF $T_{\alpha}(t)$ is constructed by a product of $\{ e^{-i a_j(t) H_j}\}$, where $e^{-i a_j(t) H_j}$ can be implemented by native quantum gates in quantum computing platforms. For example, the first- and second-order Trotter PFs, $T_1(t)$ and $T_2(t)$, are given by 
\begin{align}\label{eq:trot_1st}
T_1(t) &= \prod_{j=1}^N e^{-it H_j}=U(t)+\mathcal{O}(t^{2}),\\
T_2(t)& = T_1\left(\frac{t}{2}\right)T^\dagger_1\left(-\frac{t}{2}\right)=U(t)+\mathcal{O}(t^3), 
\end{align}
where $T^\dagger_\alpha(t) T_\alpha(t) = \mathbb{1}$ and the associated errors are called Trotter errors. We refer to a PF as symmetric if $T^\dagger_\alpha(-t) = T_\alpha(t)$, otherwise it is referred to as a regular PF, such as $T_1(t)$ is a regular PF and $T_2(t)$ is a symmetric PF.

To enhance the accuracy of quantum simulation, one may construct higher-order PFs such as~\cite{suzuki1991general, berry2007efficient},
\begin{align}
T_{2k+2} (t) = T^2_{2k}(\gamma_k t)T_{2k}(\left(1-4\gamma_k \right)t)T^2_{2k}(\gamma_k t),
\end{align}
where $\gamma^{-1}_k =  4-4^{1/(2k+1)}$ $(k \in \mathbb{Z}_{+})$, or  implement its $n$-folded PF,
 \begin{align}
 T_\alpha ^n \left(\frac{t}{n}\right) = U(t) +\mathcal{O}\left(\frac{t^{\alpha+1}}{n^\alpha}\right). \nonumber
 \end{align}
However, the higher-order PFs require exponentially long circuit depth due to their recursive nature, and $n$-fold Trotter PFs do not improve the inherent Trotter error scaling of $t$ despite demanding circuit depth that grows linearly with $n$. This requirement of long sequences of quantum gates severely ruins the outcomes of quantum simulation due to physical noise.  
 
For efficient and precise quantum simulations, the multi-product formula (MPF) approach has recently been widely adopted to reduce algorithmic error of PF-based approaches \cite{qsim_algo_LCU,blanes1999extrapolation, chin2010multi}.
The MPF approach uses a linear combination of various folded PFs to achieve desired accuracy of the simulation. In detail, for a given PF $T_\alpha(t)$ and its $n_{1,\cdots, K}$-folded PFs,  their linear combination $\sum_{i=1}^K c_i T_\alpha^{n_i}\left(\frac{t}{n_i}\right)$  where the Vandermonde equation determines coefficients~\cite{chin2010multi}, shows improved Trotter error scaling
 \begin{equation}\label{eq:mpf_basic}
\sum_{i=1}^K c_i T_\alpha^{n_i}\left(\frac{t}{n_i}\right)  = U(t) +\mathcal{O}(t^{o_{K}}),
\end{equation}
where
\begin{align}
o_K=
\begin{cases}
 \alpha + K ~~&\text{for regular PFs,}\\
 \alpha + 2K-1 ~~&\text{for symmetric PFs.}
 \label{eq:mitigation-power}
\end{cases}
\end{align} 
Furthermore, it has also been shown that if the MPF coefficients $\{c_i\}$ and folding numbers (Trotter exponents) $\{n_i\}$ satisfy Eq.~\eqref{eq:mpf_basic}, then the weighted sum of expectation values of an observable $O$ shows the same algorithmic error mitigation \cite{vazquez2023well}
\begin{align}
\langle O(t) \rangle_{\text{MPF}} &=\sum_{i=1}^K c_i \langle O(t)\rangle_{T_{\alpha}^{n_i}},\nonumber\\&= \left\langle O(t)\right\rangle_{U} +\mathcal{O}(t^{o_K}),\label{eq:mpf_extra}
\end{align}
where 
\begin{align*}
\langle O(t)\rangle_{T_\alpha^{n}}&:= \left\langle \psi \Big| {T_\alpha^\dagger}^{n}\left(\frac{t}{n}\right) O T_\alpha^{n}\left(\frac{t}{n}\right)\Big|\psi \right\rangle,\\
\langle O(t)\rangle_U&:= \langle\psi \big| U^\dagger(t) O U(t)\big| \psi\rangle,
\end{align*}
for an initial state $|\psi\rangle$. 

The MPF method for algorithmic error mitigation in Eq.~\eqref{eq:mpf_extra} is relatively easy to implement on a real quantum simulator, as it requires only repeated Trotter circuits together with classical post-processing of measurement outcomes, rather than multiple controlled-unitary gates used in cutting-edge quantum algorithms. This relatively short depth quantum circuit suppresses physical errors occurring in quantum simulations, which would otherwise exponentially deteriorate outcomes of quantum simulation as the circuit depth increases. Thus, the MPF approach is well suited for near-term quantum devices with physical noise, in contrast to other quantum simulation algorithms such as quantum singular value transformation, which typically require substantially deeper and more resource-intensive circuits~\cite{qsim_algo_LCU,qsim_algo_LCU2,qsim_algo_QSP,qsim_algo_qubitization,qsim_algo_quantumwalk,qsim_algo_quantumwalk2,qsim_algo_quantumwalk3,qsim_algo_qwalk,qsim_algo_random,qsim_algo_sym}.

\section{Dual channel MPF protocol}\label{sec:dcmpf}

While the MPF method provides a favorable way to suppress Trotter errors to arbitrary order that grows linearly with $K$, it also suffers from the so-called `conditioning problem' \cite{low2019well, vazquez2023well} and which makes the outcome from the protocol sensitive to hardware noise. Achieving a well-conditioned MPF typically requires a large folding number, resulting in substantial circuit depth, which in turn accumulates more physical error.

Here, we propose a post-processing protocol, the dual-channel multi-product formula (DCMPF) protocol. Under this protocol, Trotter error is mitigated up to $\mathcal{O}(t^{\alpha + 2K})$ with the {\it regular} PFs (c.f. Eq.~\eqref{eq:mitigation-power}), which immediately implies that our protocol is less resource-intensive and relatively stable in terms of the conditioning problem, which will be clear in a moment. 

To introduce our protocol, let us denote $\bar{T}_\alpha(t) = T_\alpha^\dagger (-t)$ for a given regular PF $T_\alpha(t)$. Note that $\bar{T}_\alpha(t)$ is a PF constructed by reversing the order of the sequence in $T_\alpha(t)$; for example, $\bar{T}_1(t)= \prod_{j=1}^N e^{-iH_{N-j+1}t}$ as obtained from the $T_1(t)$ presented in Eq.~\eqref{eq:trot_1st}. Leveraging both $T_\alpha(t)$ and $\bar{T}_\alpha(t)$, together with their folded circuits, our protocol improves the Trotter error scalings from $\mathcal{O}(t^{\alpha+K})$ to $\mathcal{O}(t^{\alpha+2K})$. 

We show that a linear combination 
{\small
\begin{align}
\frac{1}{2}\sum_{i=1}^K c_i\left(T_\alpha^{n_i}\left(\frac{t}{n_i}\right)+\bar{T}_\alpha^{n_i}\left(\frac{t}{n_i}\right)\right)=U(t)+\mathcal{O}(t^{\alpha+2K}),
\label{eq:mpf_dual_basic}
\end{align}}with coefficients $\{c_i\}$ satisfying the Vandermonde equation
{\footnotesize
\begin{equation}\label{eq:vander_dual}
\begin{bmatrix}
1&1&\cdots&1\\n_1^{-(\alpha+1)}&n_2^{-(\alpha+1)}&\cdots&n_K^{-(\alpha+1)}\\\vdots&\vdots&\ddots&\vdots\\n_1^{-2K+\alpha-1}&n_2^{-2K+\alpha-1}&\cdots&n_K^{-2K+\alpha-1}\end{bmatrix}\begin{bmatrix}c_1\\c_2\\\vdots\\c_K\end{bmatrix}=\begin{bmatrix}1\\0\\\vdots\\0\end{bmatrix},
\end{equation}
}is identical to the equation for the MPF based on symmetric PF. Furthermore, our protocol also works for expectation values of an observable $O$; the mitigated result is
\begin{align}
\langle O (t)\rangle_{\small\text{DCMPF}}&=\frac{1}{2}\sum_{i=1}^K c_i\left(\left\langle O(t)\right\rangle_{T^{n_i}_\alpha}+\left\langle O(t)\right\rangle_{\bar{T}^{n_i}_\alpha}\right), \nonumber\\
&=\langle O(t)\rangle_U +\mathcal{O}(t^{\alpha+2K}).\label{eq:mpf_dual_extra}
\end{align} 

Let us briefly outline the underlying mechanism of our proposal. For a regular PF $T_\alpha(t)$, the sum $T_\alpha(t)+\bar{T}_\alpha(t)$ satisfies the symmetric condition
\begin{align}
\left( T_\alpha(-t) + \bar{T} _\alpha (-t) \right)^{\dagger} = \left( T_\alpha (t) + \bar{T} _\alpha (t) \right),
\end{align}
and this property also valid for any $n$-fold construction. Consequently, the Trotter errors associated with the summation should behave as symmetric PFs, which allows for the cancellation of the intertwined Trotter errors across different foldings. The details of the proof are provided in Appendix~\ref{pfDCMPF}. 

We remark that DCMPF shares the same condition number \cite{low2019well, vazquez2023well} as the MPF method based on symmetric PFs, due to the same Vandermonde equation, but it requires nearly half the circuit depth to achieve the same algorithmic error. This does not merely provide a factor-of-two reduction in physical noise for the same condition number. For a given maximum circuit-depth constraint, DCMPF allows a folding number that is almost twice as large as that of the MPF based on symmetric PFs. As a result, DCMPF possesses a much larger set of folding numbers that yield a well-conditioned MPF, enabling larger values of $K$. This, in turn, leads to a notable enhancement of accuracy under the same maximum circuit-depth limitation, which will be presented in detail in Section~\ref{sec:numerics}.

As a final remark of this section, since DCMPF uses a linear combination of measurement outcomes as in Eq.~\eqref{eq:mpf_dual_extra}, the variance of the mitigated result using a given regular PF is
\begin{align*}
&\text{var} \langle O (t)\rangle_{\rm DCMPF}, \\
=& \frac{1}{4} \sum_{i=1}^{K}  c_i^2 \left( \text{var} \langle O(t) \rangle_{T_{\alpha}^{n_i}}+\text{var} \langle O(t) \rangle_{\bar{T}_{\alpha}^{n_i}}\right). 
\end{align*}
The variance of the mitigated result by DCMPF is therefore equal to the conventional MPF approach with single PFs. Consequently, a shot-cost ratio required to achieve a given algorithmic precision satisfies
\begin{equation*}
\frac{\left( \text{var}\langle O (t)\rangle_{\text{DCMPF}}/\epsilon_{\text{DCMPF}}^2 \right)}{ \left( \text{var}\langle O (t)\rangle_{\text{MPF}}/\epsilon_{\text{MPF}}^2 \right)} \propto t^{-2K},
\end{equation*}
where  $\epsilon_{\text{DCMPF,MPF}}$ denotes the precisions of the corresponding mitigation protocols.

\section{Numerical demonstration}\label{sec:numerics}
\subsection{Theoretical performance : 1D Transverse Ising model}
We first examine the $2K$ scaling of the error given in Eq.~\eqref{eq:mpf_dual_extra}. Our method is tested on the transverse-field Ising chain, defined by the Hamiltonian
\begin{equation*}
H=-J\sum_{j=1}^{N-1}\sigma_{j}^z\sigma_{j+1}^z+h\sum_{j=1}^N \sigma_j^x,
\end{equation*}
for the $N=8$ case. We set $J=1$ so that energies are measured in units of $J$, choose $h=0.5$, and compute the expectation value of $\frac{1}{N}\sum_{j=1}^N \sigma_j^z$ at time $t=0.8$, using a random product state as the initial state. We simulated with the first-order (Lie–Trotter) and the third-order (Ruth) PFs \cite{ruth1983canonical} to construct the DCMPF, and compare the results with those from the MPF method using the first-order PF. 

Because the error depends not only on $t$ but also on the folding numbers $\{n_i\}$, we choose $K$-folding numbers for each odd $K$, centered around $n_{\rm mid}=4$ with displacement 1: 
\begin{equation*}
\Big\{n_{\rm mid}-\frac{K-1}{2},\cdots,n_{\rm mid},\cdots,n_{\rm mid}+\frac{K-1}{2}\Big\},
\end{equation*}
and estimate how the error exponent depends on $K$ by fitting the data for $K = 1,3,5,7$ to the function $c_1(t/n_{\rm mid})^{c_2 K}$.  
\begin{figure}[h]
\centering
  \includegraphics[width=8cm]{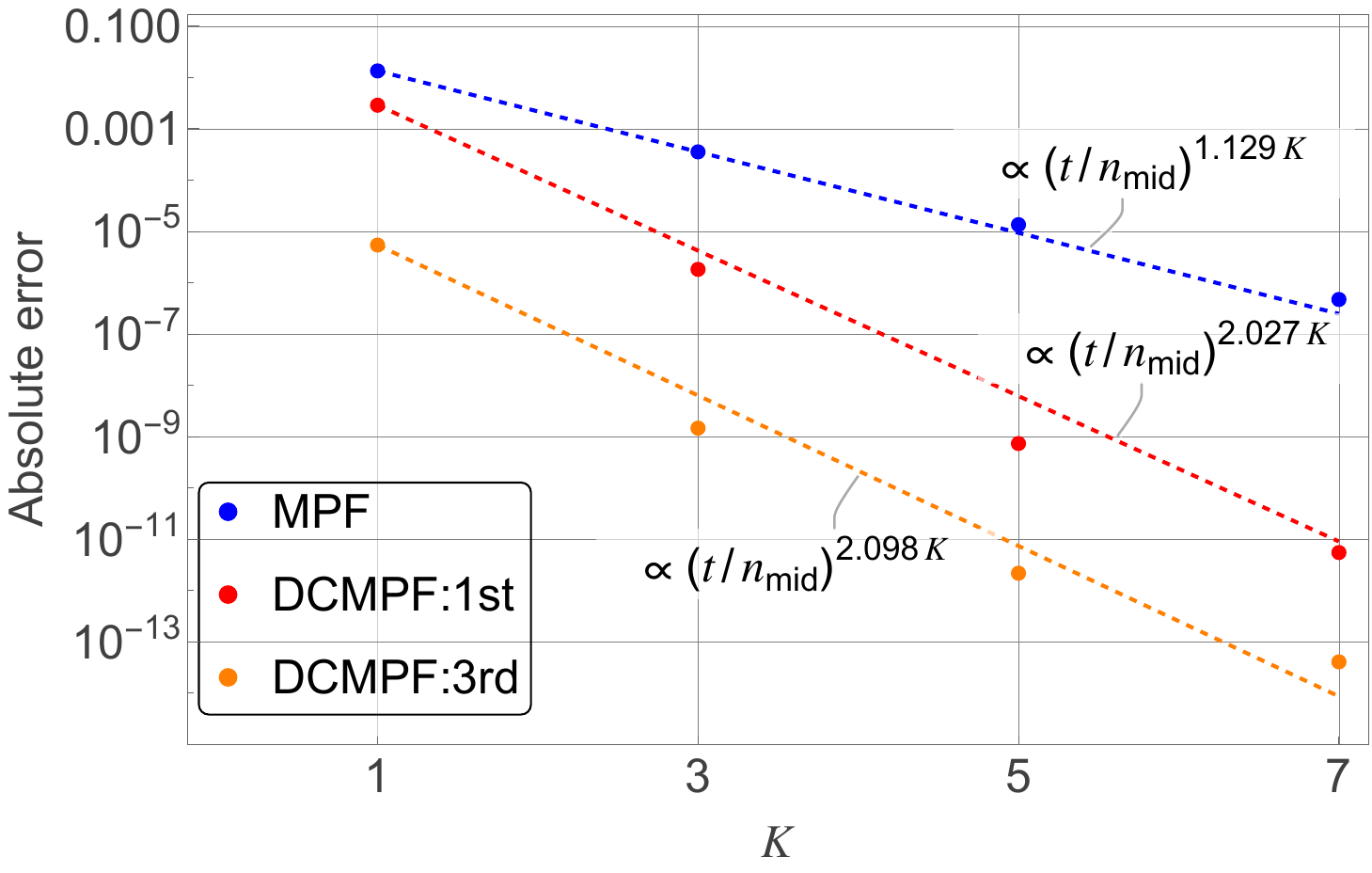}
\caption{\label{fig:fitting}  Absolute error of the global $z$-magnetization simulated using the MPF based on regular first-order PFs (blue dots), the DCMPF based on first-order PFs (red dots), and the DCMPF based on third-order PFs (orange dots), shown with respect to the exact value. The corresponding fitting results are plotted as dashed lines.}
\end{figure}

The exponents $c_2$ obtained from the fitting are approximately $1.129, 2.027$, and $2.098$ for the MPF method using the regular first-order PFs, and for the DCMPF based on the first- and third-order PFs, respectively [Fig.~\ref{fig:fitting}], and the results are in good agreement with Eq.~\eqref{eq:mpf_dual_extra}.

\subsection{Noisy simulation: XXZ spin chain}
We also numerically demonstrate our method in the presence of physical noise. We simulate the XXZ spin chain with external magnetic field
{\small
\begin{align}
\nonumber H &= -J\sum_{j=1}^{N-1}\left(\sigma_j^x\sigma_{j+1}^x+\sigma_j^y\sigma_{j+1}^y+\delta\sigma_j^z\sigma_{j+1}^z\right)+h\sum_{j=1}^N\sigma_j^z, 
\end{align}
}and again we set $J=1$. For $N=8$, $\delta=-1.5$ and $h=0.1$, 
we measure the magnetization on the even sites,
\begin{equation*}
\frac{1}{4}\sum_{j=1}^4\langle\sigma_{2j}^z(t) \rangle,
\end{equation*}
at time $t=0.3$, for the N\'{e}el state $|\downarrow\uparrow\downarrow\uparrow\downarrow\uparrow\downarrow\uparrow\rangle$ as an initial state. 

\begin{figure}[h]
\centering
  \includegraphics[width=8cm]{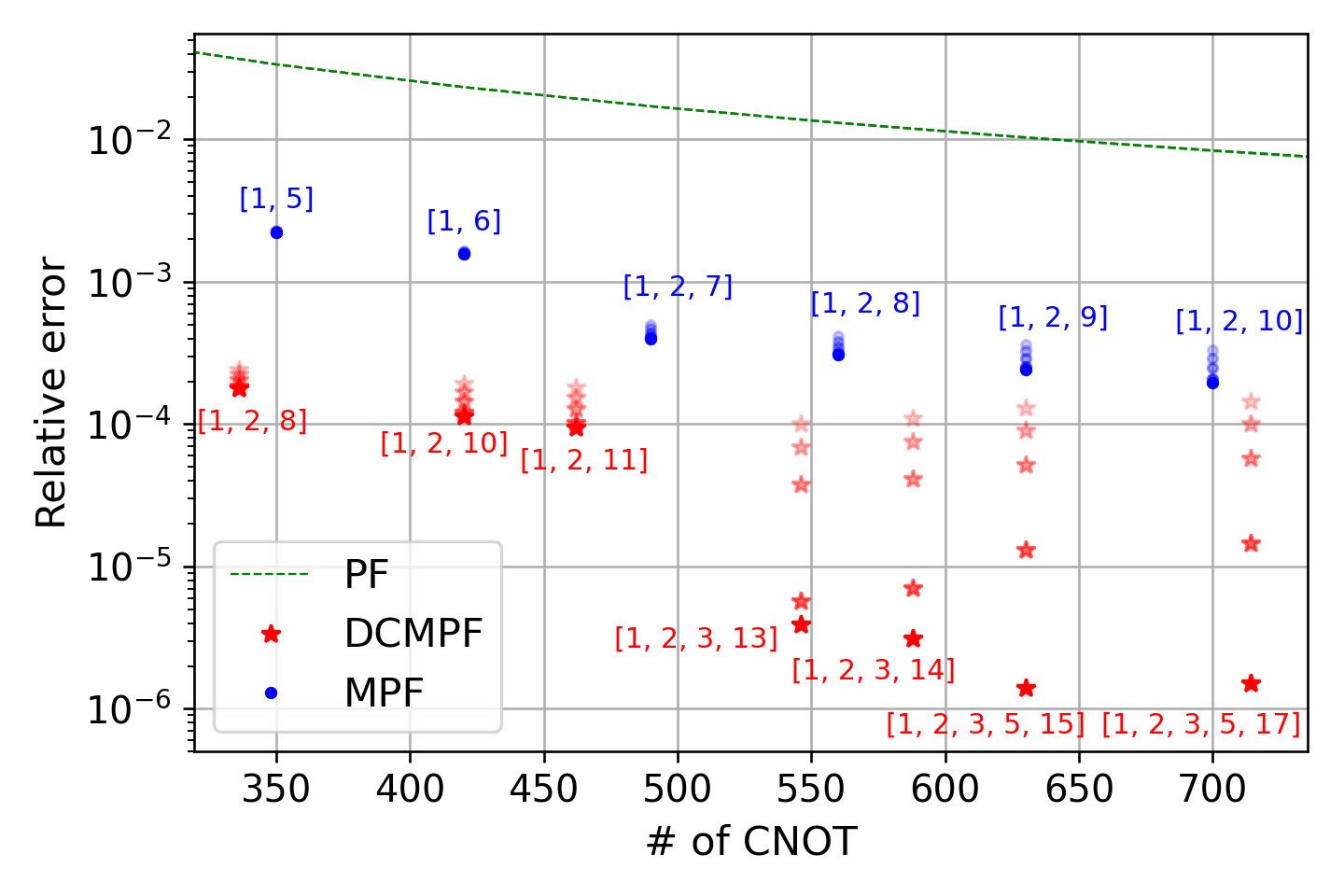}
\caption{\label{fig:numerics}  Relative error of the simulated average $z$-magnetization with respect to the exact value, plotted against the number of CNOT gates (in the deepest circuit when implemented within the MPFs). The dynamics are simulated using the second-order Suzuki-Trotter PF (green dashed lines), the MPF method based on the second-order symmetric Trotter PFs (blue dots), and our MPF method using the first-order Trotter PFs (red stars). Depolarizing noise is applied to each CNOT gate with probabilities $10^{-8},10^{-7},4\times10^{-7},7\times10^{-7}$ and $10^{-6}$, corresponding to data points shown from vivid to gradually transparent colors. The folding numbers chosen for each MPF simulation are displayed together.}
\end{figure}

In order to consider the effect of physical error, we add the depolarizing noise to every CNOT gates with probabilities varying from $10^{-8}$ to $10^{-6}$.  To consider the effect of fixed resources, we plot mitigated relative errors with respect to circuit depth which measured by the number of consumed CNOT gates for the circuit. 
As a benchmark, we also plot the same relative errors mitigated by the symmetric PF [Fig. \ref{fig:numerics}].
Across fixed number of CNOT gates, we observe that DCMPF shows better performances over all tested noise levels, compared to the MPF method based on symmetric PFs, and the second-order PF.

Since we are considering a noisy environment, the conditioning problem becomes crucial~\cite{low2019well,vazquez2023well,MPF_ref_2}. To address this point, we set the criterion for a well-conditioned MPF as a condition number smaller than target values: ($||{\bf c}||_1 <1.2\sim1.3$), and choose the folding number that yields the minimum condition number.

We attempt to mimic the effect of physical noise of the noisy quantum simulator, where long sequences of quantum gates with physical noise tends to wash out the ideal measurement outcomes. Although DCMPF and MPF method based on symmetric PFs share the same condition number, the maximum achievable circuit depth differs by nearly a factor of two. Consequently, under a limited circuit depth imposed by a measure of physical constraints, our protocol admits a larger set of folding-number choices that yield well-conditioned, leading to higher accuracy compared to the MPF method [Fig. \ref{fig:numerics}].

\section{Conclusion} \label{sec:conclusion}
In this work, we propose the dual channel multi-product formula (DCMPF) protocol, exploiting a PF and its inverse circuit to improve the performance of Trotter error mitigation, which effectively halves the quantum circuit depth compared to the MPF method with symmetric PFs. These structural features may directly influence both physical noise and shot noise, and become especially advantageous when circuit resources are limited. This half-depth property often yields far greater practical benefits than the naive linear reduction might suggest. A concrete and illustrative advantage of the circuit depth reduction is the enlarged sets of foldings for a well-conditioned MPF \cite{low2019well, vazquez2023well}.

When it comes to the implementation of our protocol in the quantum simulator, physical error should be considered. To mitigate the effect of physical error, even though we are not directly mitigating physical error, the conditioning problem under the fixed circuit depth should be important. Considering this point, we numerically support that our proposal shows remarkably better algorithmic mitigation performances. This improvement came from the shorter circuit depth with the same variance, since without any physical error mitigation, the physical errors in quantum circuits exponentially degrade the quality of outcomes as a function of circuit depth.

Furthermore, we believe that an algorithmic error-mitigation method with short circuit depth may also reduce the overhead required for physical error mitigation. Thus, for future work, it is promising to develop a physical error-mitigation technique suitable for our proposal and to implement it on a quantum simulator.

\section{Acknowledgement}
This work was supported by the National Research Foundation of Korea(NRF) grant funded by the Korea government(MSIT) (RS-2023-NR119931, RS-2025-18362970), the Korean ARPA-H Project through the Korea Health Industry Development Institute (KHIDI) funded by the Ministry of Health \& Welfare, Republic of Korea (RS-2025-25456722), and the Ministry of Trade, Industry, and Energy (MOTIE, Korea, under the project “Industrial Technology Infrastructure Program” (RS2024-00466693).
SJL is funded by KIAS Individual Grant via the Quantum Universe Center (QP104301-6P104301).

\paragraph{Data availability statement: }All data and codes in this work are available from the corresponding authors upon reasonable request.

\bibliographystyle{unsrt}
\bibliography{mybibfile}

@article{Miessen2023_qs_review,
	author = {Miessen, Alexander and Ollitrault, Pauline J. and Tacchino, Francesco and Tavernelli, Ivano},
	date = {2023/01/01},
	doi = {10.1038/s43588-022-00374-2},
	id = {Miessen2023},
	isbn = {2662-8457},
	journal = {Nature Computational Science},
	number = {1},
	pages = {25--37},
	title = {Quantum algorithms for quantum dynamics},
	url = {https://doi.org/10.1038/s43588-022-00374-2},
	volume = {3},
	year = {2023}}

@article{Georgescu2014_qs_review,
  title = {Quantum simulation},
  author = {Georgescu, I. M. and Ashhab, S. and Nori, Franco},
  journal = {Rev. Mod. Phys.},
  volume = {86},
  issue = {1},
  pages = {153--185},
  numpages = {33},
  year = {2014},
  month = {Mar},
  publisher = {American Physical Society},
  doi = {10.1103/RevModPhys.86.153},
  url = {https://link.aps.org/doi/10.1103/RevModPhys.86.153}
}

@article{Martyn2021_gradn_unification,
  title = {Grand Unification of Quantum Algorithms},
  author = {Martyn, John M. and Rossi, Zane M. and Tan, Andrew K. and Chuang, Isaac L.},
  journal = {PRX Quantum},
  volume = {2},
  issue = {4},
  pages = {040203},
  numpages = {40},
  year = {2021},
  month = {Dec},
  publisher = {American Physical Society},
  doi = {10.1103/PRXQuantum.2.040203},
  url = {https://link.aps.org/doi/10.1103/PRXQuantum.2.040203}
}

@article{Childs2021,
  title = {Theory of Trotter Error with Commutator Scaling},
  author = {Childs, Andrew M. and Su, Yuan and Tran, Minh C. and Wiebe, Nathan and Zhu, Shuchen},
  journal = {Phys. Rev. X},
  volume = {11},
  issue = {1},
  pages = {011020},
  numpages = {49},
  year = {2021},
  month = {Feb},
  publisher = {American Physical Society},
  doi = {10.1103/PhysRevX.11.011020},
  url = {https://link.aps.org/doi/10.1103/PhysRevX.11.011020}
}

@article{PF_nonsymmetric,
	author = {Morales, Mauro E. S. and Costa, Pedro C. S. and Pantaleoni, Giacomo and Burgarth, Daniel K. and Sanders, Yuval R. and Berry, Dominic W.},
	doi = {10.2478/qic-2025-0001},
	journal = {Quantum Information \& Computation},
	number = {1},
	pages = {1--35},
	title = {Selection and Improvement of Product Formulae for Best Performance of Quantum Simulation},
	url = {https://doi.org/10.2478/qic-2025-0001},
	volume = {25},
	year = {2025}}

@misc{TEP,
	archiveprefix = {arXiv},
	author = {Sangjin Lee and Youngseok Kim and Seung-Woo Lee},
	eprint = {2503.09710},
	primaryclass = {quant-ph},
	title = {Trotter error mitigation by error profiling with shallow quantum circuit},
	url = {https://arxiv.org/abs/2503.09710},
	year = {2025}}

@article{qsim_algo_qwalk,
	author = {Dominic W. Berry, Andrew M. Childs},
	doi = {10.26421/qic12.1-2},
	issn = {1533-7146},
	journal = {Quantum Information and Computation},
	month = jan,
	number = {1 \& 2},
	pages = {0029-0062},
	publisher = {Rinton Press},
	title = {Black-box Hamiltonian simulation and unitary implementation},
	url = {http://dx.doi.org/10.26421/QIC12.1-2},
	volume = {12},
	year = {2012}}

@article{PF_ruth,
	author = {Ruth, Ronald D.},
	doi = {10.1109/TNS.1983.4332919},
	journal = {IEEE Transactions on Nuclear Science},
	number = {4},
	pages = {2669-2671},
	title = {A Canonical Integration Technique},
	volume = {30},
	year = {1983}}

@inbook{Suzuki2,
	address = {Berlin, Heidelberg},
	author = {Hatano, Naomichi and Suzuki, Masuo},
	booktitle = {Quantum Annealing and Other Optimization Methods},
	doi = {10.1007/11526216_2},
	editor = {Das, Arnab and K. Chakrabarti, Bikas},
	isbn = {978-3-540-31515-5},
	pages = {37--68},
	publisher = {Springer Berlin Heidelberg},
	title = {Finding Exponential Product Formulas of Higher Orders},
	url = {https://doi.org/10.1007/11526216_2},
	year = {2005}}

@article{qsim_algo_random,
	author = {Childs, Andrew M. and Ostrander, Aaron and Su, Yuan},
	doi = {10.22331/q-2019-09-02-182},
	issn = {2521-327X},
	journal = {{Quantum}},
	month = sep,
	pages = {182},
	publisher = {{Verein zur F{\"{o}}rderung des Open Access Publizierens in den Quantenwissenschaften}},
	title = {Faster quantum simulation by randomization},
	url = {https://doi.org/10.22331/q-2019-09-02-182},
	volume = {3},
	year = {2019}}

@article{qsim_algo_sym,
	author = {Tran, Minh C. and Su, Yuan and Carney, Daniel and Taylor, Jacob M.},
	doi = {10.1103/PRXQuantum.2.010323},
	issue = {1},
	journal = {PRX Quantum},
	month = {Feb},
	numpages = {28},
	pages = {010323},
	publisher = {American Physical Society},
	title = {Faster Digital Quantum Simulation by Symmetry Protection},
	url = {https://link.aps.org/doi/10.1103/PRXQuantum.2.010323},
	volume = {2},
	year = {2021}}

@misc{MPF_ref_4,
	archiveprefix = {arXiv},
	author = {James D. Watson and Jacob Watkins},
	eprint = {2408.14385},
	primaryclass = {quant-ph},
	title = {Exponentially Reduced Circuit Depths Using Trotter Error Mitigation},
	url = {https://arxiv.org/abs/2408.14385},
	year = {2024}}

@inproceedings{qsim_algo_quantumwalk2,
	address = {New York, NY, USA},
	author = {Childs, Andrew M. and Cleve, Richard and Deotto, Enrico and Farhi, Edward and Gutmann, Sam and Spielman, Daniel A.},
	booktitle = {Proceedings of the Thirty-Fifth Annual ACM Symposium on Theory of Computing},
	doi = {10.1145/780542.780552},
	isbn = {1581136749},
	location = {San Diego, CA, USA},
	numpages = {10},
	pages = {59--68},
	publisher = {Association for Computing Machinery},
	series = {STOC '03},
	title = {Exponential algorithmic speedup by a quantum walk},
	url = {https://doi.org/10.1145/780542.780552},
	year = {2003}}

@article{MPF_extrapol,
	author = {Rendon, Gumaro and Watkins, Jacob and Wiebe, Nathan},
	doi = {10.22331/q-2024-02-26-1266},
	issn = {2521-327X},
	journal = {{Quantum}},
	month = feb,
	pages = {1266},
	publisher = {{Verein zur F{\"{o}}rderung des Open Access Publizierens in den Quantenwissenschaften}},
	title = {Improved {A}ccuracy for {T}rotter {S}imulations {U}sing {C}hebyshev {I}nterpolation},
	url = {https://doi.org/10.22331/q-2024-02-26-1266},
	volume = {8},
	year = {2024}}

@article{qsim_algo_LCU,
	author = {Andrew M. Childs, Nathan Wiebe},
	doi = {10.26421/qic12.11-12},
	issn = {1533-7146},
	journal = {Quantum Information and Computation},
	month = nov,
	number = {11 \& 12},
	pages = {0901-0924},
	publisher = {Rinton Press},
	title = {Hamiltonian Simulation Using Linear Combinations of Unitary Operations},
	url = {http://dx.doi.org/10.26421/QIC12.11-12},
	volume = {12},
	year = {2012}}

@article{qsim_algo_qubitization,
	author = {Low, Guang Hao and Chuang, Isaac L.},
	doi = {10.22331/q-2019-07-12-163},
	issn = {2521-327X},
	journal = {{Quantum}},
	month = jul,
	pages = {163},
	publisher = {{Verein zur F{\"{o}}rderung des Open Access Publizierens in den Quantenwissenschaften}},
	title = {Hamiltonian {S}imulation by {Q}ubitization},
	url = {https://doi.org/10.22331/q-2019-07-12-163},
	volume = {3},
	year = {2019}}

@article{qsim_general,
	author = {Cirac, J. Ignacio and Zoller, Peter},
	date = {2012/04/01},
	doi = {10.1038/nphys2275},
	id = {Cirac2012},
	isbn = {1745-2481},
	journal = {Nature Physics},
	number = {4},
	pages = {264--266},
	title = {Goals and opportunities in quantum simulation},
	url = {https://doi.org/10.1038/nphys2275},
	volume = {8},
	year = {2012}}

@article{PF_ruth2,
	author = {Etienne Forest and Ronald D. Ruth},
	doi = {https://doi.org/10.1016/0167-2789(90)90019-L},
	issn = {0167-2789},
	journal = {Physica D: Nonlinear Phenomena},
	number = {1},
	pages = {105-117},
	title = {Fourth-order symplectic integration},
	url = {https://www.sciencedirect.com/science/article/pii/016727899090019L},
	volume = {43},
	year = {1990}}

@article{qsim_algo_QSP,
	author = {Low, Guang Hao and Chuang, Isaac L.},
	doi = {10.1103/PhysRevLett.118.010501},
	issue = {1},
	journal = {Phys. Rev. Lett.},
	month = {Jan},
	numpages = {5},
	pages = {010501},
	publisher = {American Physical Society},
	title = {Optimal Hamiltonian Simulation by Quantum Signal Processing},
	url = {https://link.aps.org/doi/10.1103/PhysRevLett.118.010501},
	volume = {118},
	year = {2017}}

@article{qsim_algo_quantumwalk3,
	author = {Childs, Andrew M.},
	date = {2010/03/01},
	doi = {10.1007/s00220-009-0930-1},
	id = {Childs2010},
	isbn = {1432-0916},
	journal = {Communications in Mathematical Physics},
	number = {2},
	pages = {581--603},
	title = {On the Relationship Between Continuous- and Discrete-Time Quantum Walk},
	url = {https://doi.org/10.1007/s00220-009-0930-1},
	volume = {294},
	year = {2010}}

@misc{MPF_ref_3,
	archiveprefix = {arXiv},
	author = {Junaid Aftab and Dong An and Konstantina Trivisa},
	eprint = {2403.08922},
	primaryclass = {quant-ph},
	title = {Multi-product Hamiltonian simulation with explicit commutator scaling},
	url = {https://arxiv.org/abs/2403.08922},
	year = {2024}}

@article{qsim_algo_quantumwalk,
	author = {Berry, Dominic W. and Childs, Andrew M. and Ostrander, Aaron and Wang, Guoming},
	date = {2017/12/01},
	doi = {10.1007/s00220-017-3002-y},
	id = {Berry2017},
	isbn = {1432-0916},
	journal = {Communications in Mathematical Physics},
	number = {3},
	pages = {1057--1081},
	title = {Quantum Algorithm for Linear Differential Equations with Exponentially Improved Dependence on Precision},
	url = {https://doi.org/10.1007/s00220-017-3002-y},
	volume = {356},
	year = {2017}}

@article{qsim_general3,
	author = {Monroe, C. and Campbell, W. C. and Duan, L.-M. and Gong, Z.-X. and Gorshkov, A. V. and Hess, P. W. and Islam, R. and Kim, K. and Linke, N. M. and Pagano, G. and Richerme, P. and Senko, C. and Yao, N. Y.},
	doi = {10.1103/RevModPhys.93.025001},
	issue = {2},
	journal = {Rev. Mod. Phys.},
	month = {Apr},
	numpages = {57},
	pages = {025001},
	publisher = {American Physical Society},
	title = {Programmable quantum simulations of spin systems with trapped ions},
	url = {https://link.aps.org/doi/10.1103/RevModPhys.93.025001},
	volume = {93},
	year = {2021}}

@article{qsim_algo_LCU2,
	author = {Berry, Dominic W. and Childs, Andrew M. and Cleve, Richard and Kothari, Robin and Somma, Rolando D.},
	doi = {10.1103/PhysRevLett.114.090502},
	issue = {9},
	journal = {Phys. Rev. Lett.},
	month = {Mar},
	numpages = {5},
	pages = {090502},
	publisher = {American Physical Society},
	title = {Simulating Hamiltonian Dynamics with a Truncated Taylor Series},
	url = {https://link.aps.org/doi/10.1103/PhysRevLett.114.090502},
	volume = {114},
	year = {2015}}

@article{MPF_ref_2,
	author = {Zhuk, Sergiy and Robertson, Niall F. and Bravyi, Sergey},
	doi = {10.1103/PhysRevResearch.6.033309},
	issue = {3},
	journal = {Phys. Rev. Res.},
	month = {Sep},
	numpages = {19},
	pages = {033309},
	publisher = {American Physical Society},
	title = {Trotter error bounds and dynamic multi-product formulas for Hamiltonian simulation},
	url = {https://link.aps.org/doi/10.1103/PhysRevResearch.6.033309},
	volume = {6},
	year = {2024}}

@article{qsim_appl2,
	author = {B. P. Lanyon and C. Hempel and D. Nigg and M. M{\"u}ller and R. Gerritsma and F. Z{\"a}hringer and P. Schindler and J. T. Barreiro and M. Rambach and G. Kirchmair and M. Hennrich and P. Zoller and R. Blatt and C. F. Roos},
	doi = {10.1126/science.1208001},
	journal = {Science},
	number = {6052},
	pages = {57-61},
	title = {Universal Digital Quantum Simulation with Trapped Ions},
	url = {https://www.science.org/doi/abs/10.1126/science.1208001},
	volume = {334},
	year = {2011}}

@article{lloyd1996universal,
	author = {Lloyd, Seth},
	journal = {Science},
	number = {5278},
	pages = {1073--1078},
	publisher = {American Association for the Advancement of Science},
	title = {Universal quantum simulators},
	volume = {273},
	year = {1996}}

@article{berry2007efficient,
	author = {Berry, Dominic W and Ahokas, Graeme and Cleve, Richard and Sanders, Barry C},
	journal = {Communications in Mathematical Physics},
	number = {2},
	pages = {359--371},
	publisher = {Springer},
	title = {Efficient quantum algorithms for simulating sparse Hamiltonians},
	volume = {270},
	year = {2007}}

@article{suzuki1991general,
	author = {Suzuki, Masuo},
	journal = {Journal of mathematical physics},
	number = {2},
	pages = {400--407},
	publisher = {American Institute of Physics},
	title = {General theory of fractal path integrals with applications to many-body theories and statistical physics},
	volume = {32},
	year = {1991}}

@article{chin2010multi,
	author = {Chin, Siu A},
	journal = {Celestial Mechanics and Dynamical Astronomy},
	number = {4},
	pages = {391--406},
	publisher = {Springer},
	title = {Multi-product splitting and Runge-Kutta-Nystr{\"o}m integrators},
	volume = {106},
	year = {2010}}

@article{blanes1999extrapolation,
	author = {Blanes, S and Casas, F and Ros, J},
	journal = {Celestial Mechanics and Dynamical Astronomy},
	number = {2},
	pages = {149--161},
	publisher = {Springer},
	title = {Extrapolation of symplectic integrators},
	volume = {75},
	year = {1999}}

@article{vazquez2023well,
	author = {Vazquez, Almudena Carrera and Egger, Daniel J and Ochsner, David and Woerner, Stefan},
	journal = {Quantum},
	pages = {1067},
	publisher = {Verein zur F{\"o}rderung des Open Access Publizierens in den Quantenwissenschaften},
	title = {Well-conditioned multi-product formulas for hardware-friendly Hamiltonian simulation},
	volume = {7},
	year = {2023}}

@article{low2019well,
	author = {Low, Guang Hao and Kliuchnikov, Vadym and Wiebe, Nathan},
	journal = {arXiv preprint arXiv:1907.11679},
	title = {Well-conditioned multiproduct Hamiltonian simulation},
	year = {2019}}

@article{ruth1983canonical,
  title={A canonical integration technique},
  author={Ruth, Ronald D},
  journal={IEEE Trans. Nucl. Sci.},
  volume={30},
  number={CERN-LEP-TH-83-14},
  pages={2669--2671},
  year={1983}
}

@article{childs2019nearly,
  title={Nearly optimal lattice simulation by product formulas},
  author={Childs, Andrew M and Su, Yuan},
  journal={Physical review letters},
  volume={123},
  number={5},
  pages={050503},
  year={2019},
  publisher={APS}
}

\onecolumn\newpage
\appendix

\setcounter{table}{0}  
  \renewcommand{\thetable}{A\arabic{table}} 
  \setcounter{figure}{0} 
  \renewcommand{\thefigure}{A\arabic{figure}}
  \setcounter{equation}{0} 
  \renewcommand{\theequation}{A\arabic{equation}}

\section{Detailed calculation of conventional MPF}

To be self-contained, we provide details of the calculation of the conventional MPF approach, which helps to follow the logic of our dual channel MPF. 
\subsection{Conventional MPF: operator level, Eq.~\eqref{eq:mpf_basic}}
For any given $\alpha$-order Trotter PF $T_\alpha(t)$, one can write the formula in terms of Trotter error operators as 
\begin{align*}
T_\alpha(t) &= e^{-iHt +i\sum_{q=\alpha+1}^\infty t^qE^{(0)}_q},\\
&= e^{-iHt}+\sum_{q=\alpha+1}^\infty t^q E_q,
\end{align*}
where $E^{(0)}_q$ is Hermitian for all $q$ due to the unitarity of  $T_\alpha$. While the details of Trotter errors $E^{(0)}_q$ and $E_q$ are determined by the specific structure of the PF, the relation between $E^{(0)}_q$ and $E_q$ can be derived independently of that structure, by using

\begin{equation}\label{eq:temp}
\sum_{q=\alpha+1}^\infty t^q E_q ={\rm exp}[itH]{\rm exp}\left[-itH +i\sum_{q=\alpha+1}^\infty t^q E^{(0)}_q\right]-1.
\end{equation}
By comparing the terms at each order in $t$, we can write the $E_{q}$ in terms of $E^{(0)}_q$ as 
\begin{equation}\label{eq:c_func}
{E}_{q}=\sum_{l,l'}C^{(q)}_{l,l'}(H,E^{(0)}_{q},E^{(0)}_{q-1},\cdots,E^{(0)}_{\alpha+1}).
\end{equation}
The function $C^{(q)}_{l,l'}$ is a sum of terms composed of $l$ number of $H$, and $l'$ number of error operators $E^{(0)}$, under the constraint that the sum of the order of all including error operators and the number of $H$ is equivalent to $q$. For example, explicit forms of some $C^{(q)}$ are
\begin{equation*} 
C^{(q)}_{0,1}=iE^{(0)}_{q}\;,\;\;C^{(q)}_{1,1}=-\frac{1}{2}[H,E^{(0)}_{q-1}]+HE^{(0)}_{q-1}\;,\;\;C^{(q)}_{0,2}=-\frac{1}{2}\sum_{q'=\alpha+1}^{q-(\alpha+1)}E^{(0)}_{q'}E^{(0)}_{q-q'}\;,\;\;\cdots.
\end{equation*}
We note that this expression is useful when we look into the error of $n$-fold Trotter PF. Similarly, the $n$-fold Trotter PF can be written as
\begin{align}\label{eq:e0_nq}
T_\alpha^n\left(\frac{t}{n}\right)&=e^{-itH+i\sum_{q=\alpha+1}^\infty t^q E^{(0)}_{n;q}},\\
&=e^{-itH}+\sum_{q=\alpha+1}^\infty t^q E_{n;q},
\end{align}
where $E^{(0)}_{n;q}=\frac{E^{(0)}_q}{n^{q-1}}$. 
In this expression, the relation between $E^{(0)}_{n;q}$ and $E_{n;q}$ is the same with those of $E^{(0)}_q$ and $E_q$, because the function $C_{l,l'}$ is determined independently of the specific structure of the PF. Now we can rewrite the $E_{n;q}$ in terms of $E^{(0)}_q$:  
\begin{align}
{E}_{n;q}&=\sum_{l,l'}C^{(q)}_{l,l'}(H,E^{(0)}_{n;q},E^{(0)}_{n;q-1},\cdots,E^{(0)}_{n;\alpha+1}),\\&=\sum_{l,l'}\frac{1}{n^{q-(l+l')}}C^{(q)}_{l,l'}(H,E^{(0)}_{q},E^{(0)}_{q-1},\cdots,E^{(0)}_{\alpha+1}),\label{eq:e_q_to_e_nq}
\end{align}
and any $n_i$-fold Trotter PF is written in terms of $E^{(0)}_q$ as
\begin{align*}
T^{n_i}_\alpha\left(\frac{t}{n_i}\right)&=U(t)+\sum_{q=\alpha+1}^\infty t^q E_{n_i;q},\\&=U(t)+\sum_{q=\alpha+1}^\infty t^q \sum_{l,l'}\frac{1}{n_i^{q-(l+l')}}C^{(q)}_{l,l'}(H,E^{(0)}_q,E^{(0)}_{q-1},\cdots,E^{(0)}_{\alpha+1}).
\end{align*}
The $C^{(q)}_{l,l'}$ is independent of $n_i$, and for each $t^q$, the Trotter error operator $E_{n_i;q}$ is a polynomial of $\{1/n_i^{\alpha}, 1/n_i^{\alpha+1}, \cdots, 1/n_i^{q-1}\}$, with the $n_i$-independent operator-wise coefficients. Thus, we conclude that 
\begin{equation}
\sum_{i=1}^K c_i T^{n_i}_\alpha\left(\frac{t}{n_i}\right) = U(t)+\mathcal{O}( t^{\alpha+K}),
\end{equation}
where the coefficients $\{c_k\}$ is determined by
\begin{equation}\label{eq:vander}
\begin{bmatrix}1&1&\cdots&1\\n_1^{-\alpha}&n_2^{-\alpha}&\cdots&n_K^{-\alpha}\\\vdots&\vdots&\ddots&\vdots\\n_1^{-(K-1)}&n_2^{-(K-1)}&\cdots&n_K^{-(K-1)}\end{bmatrix}\begin{bmatrix}c_1\\c_2\\\vdots\\c_K\end{bmatrix}=\begin{bmatrix}1\\0\\\vdots\\0\end{bmatrix}.
\end{equation}

The case of the symmetric PF can be proven as follows. For the $T_2(t)$, second-order symmetric Trotter formula (higher-order case can be treated straightforwardly), we can write the $n$-fold Trotter PF,
\begin{align*}
T_2^n(t/n)&=e^{-itH+i\left(t^3E^{(0)}_{n;3}+t^5 E^{(0)}_{n;5}+\cdots\right)},\\&=e^{-itH}+\sum_{q=3}^{\infty} t^q E_{n;q}.
\end{align*}
Since $E^{(0)}_{n;q}=0$ for all even $q$, it can be easily  shown that for given odd $q$, both $E_{n;q}$ and $E_{n;q+1}$ are polynomials of $1/n^2,1/n^4,\cdots,1/n^{q-1}$. 
Thus, for the second-order symmetric formula, the coefficient of the MPF satisfies
\begin{equation}\label{eq:evenvander}
\sum_{i=1}^K c_i=1\;,~\textrm{and}\;\;\sum_{i=1}^K \frac{c_i}{n_i^q}=0,~\textrm{for}\;\;q=2,4,\cdots,2(K-1),
\end{equation} 
and the error order becomes $o_K = 2+2K-1$.

\subsection{Conventional MPF: expectation value level,  Eq.~\eqref{eq:mpf_extra}}
Here, we will show that if the MPF works for individual operators as
\begin{equation}\label{eq:mpf_op}
\sum_{i=1}^K c_i T^{n_i}_\alpha\left(\frac{t}{n_i}\right)=U(t)+\mathcal{O}(t^{o_K}),
\end{equation}
and it works similarly for expectation values of an observable $O$ and states, that is 
\begin{equation}\label{eq:mpf_expec}
\sum_{i=1}^Kc_i\left\langle O(t)\right\rangle_{T^{n_i}_\alpha} =\langle O(t)\rangle_U +\mathcal{O}(t^{o_K}).
\end{equation}
The Eq.~\eqref{eq:mpf_op} implies that, in terms of $E_{n;q}$ operators, 
\begin{equation}\label{eq:simple_cond_1}
\sum_{i=1}^K c_i E_{n_i;q}=0\;,\;\;\forall q<o_K.
\end{equation} 
The expectation value $\langle O(t)\rangle_{T^{n_i}}$ has three error terms:
\begin{align*}
&\sum_{i=1}^Kc_i\left\langle O(t)\right\rangle_{T^{n_i}_\alpha}\\
=&\langle O(t)\rangle_U+ \sum_{i=1}^K\sum_{q=\alpha+1}^\infty\left(c_it^q\langle U^\dagger O E_{n_i;q}\rangle+c_it^q\langle E_{n_i;q}^\dagger O U\rangle\right) +\sum_{i=1}^K\sum_{q,q'=\alpha+1}^\infty c_it^{q+q'}\langle E_{n_i;q}^\dagger O E_{n_i;q'}\rangle.
\end{align*}
Since the first and second terms have $(o_K-1)$-order cancellation of error due to Eq.~\eqref{eq:simple_cond_1}, we will focus on the last term. 
By using the relation Eq.~\eqref{eq:e_q_to_e_nq}, the last error term for each $t^q$ order becomes 
\begin{align*}
\sum_{i=1}^K c_it^{q}\sum_{q'=\alpha+1}^{q-(\alpha+1)}\langle E_{n_i;q-q'}^\dagger O E_{n_i;q'}\rangle&=\sum_{i=1}^K c_i t^q\sum_{q'=\alpha+1}^{q-(\alpha+1)}\sum_{l,l'\bar{l},\bar{l}'}\frac{1}{n_i^{q-(l+l'+\bar{l}+\bar{l}')}}\langle (C^{(q-q')}_{l,l'})^\dagger OC^{(q')}_{\bar{l},\bar{l}'}\rangle.
\end{align*}
Thus, the error of the expectation value for each $t^q$ is a polynomial of $\{1/n_i^{\alpha},1/n_i^{\alpha+1},\cdots, 1/n_i^{q-1}\}$ with the fixed coefficient independent of $n_i$. Thus, the coefficients following Eq.~\eqref{eq:vander} also work for the expectation value formulation given in Eq.~\eqref{eq:mpf_expec}. The case of using symmetric PF can be proven in the similar manner.

\section{Proofs of dual channel MPF}\label{pfDCMPF}
In this section, we prove the performance of the dual channel MPF with detailed calculations.
\subsection{Dual channel MPF: operator level, Eq.~\eqref{eq:mpf_dual_basic}}\label{sec:proof_mpf_dual_basic}
Recall that $\bar{T}_\alpha(t) = T^\dagger_\alpha(-t)$ and the $n$-fold PF of $\bar{T}$ can be written as 
\begin{align*}
\bar{T}_\alpha^n\left(\frac{t}{n}\right)&=e^{-itH+i\sum_{q=\alpha+1}^\infty t^q \bar{E}^{(0)}_{n;q}}\\&=e^{-itH}+\sum_{q=\alpha+1}^\infty t^q \bar{E}_{n;q},
\end{align*}
where $\bar{E}^{(0)}_{n;q}=(-1)^{q-1}E^{(0)}_{n;q}$, with the $E^{(0)}_{n;q}$ in Eq.~\eqref{eq:e0_nq}. 
As in the calculation of conventional MPF, we obtain
\begin{align*}
\bar{E}_{n;q}&=\sum_{l,l'}C^{(q)}_{l,l'}(H,\bar{E}^{(0)}_{n;q},\bar{E}^{(0)}_{n;q-1},\cdots,\bar{E}^{(0)}_{n;\alpha+1}),\\&=(-1)^{q-(l+l')}\sum_{l,l'}C^{(q)}_{l,l'}(H,E^{(0)}_{n;q},E^{(0)}_{n;q-1},\cdots,E^{(0)}_{n;\alpha+1}).
\end{align*}
We can show that summed two PFs $\{T_\alpha\left( \frac{t}{n_i} \right), \bar{T}_\alpha\left( \frac{t}{n_i} \right) \}$, with the the above relation has the following structures,
\begin{align*}
T_\alpha^{n_i}\left(\frac{t}{n_i}\right)+\bar{T}_\alpha^{n_i}\left(\frac{t}{n_i}\right)=U(t)+\frac{1}{2}\sum_{q=\alpha+1}^\infty t^q \sum_{l,l'}\frac{1+(-1)^{q-(l+l')}}{n_i^{q-(l+l')}}C^{(q)}_{l,l'}(H,E^{(0)}_{q},E^{(0)}_{q-1},\cdots,E^{(0)}_{\alpha+1}).
\end{align*}
We remark that all $C^{(q)}_{l,l'}$ included in each $t^q$ have only even order coefficient, that is, $1/n_i^{\alpha+1},1/n_i^{\alpha+3},\cdots$ for $\alpha\in\mathbb{Z}_{\rm odd}$. Thus, the coefficients satisfying Eq.~\eqref{eq:vander_dual} cancel the error in order $o_K=\alpha+2K$, which completes our proof. 

\subsection{Dual channel MPF: expectation value level, Eq.~\eqref{eq:mpf_dual_extra}}\label{sec:proof_mpf_dual_extra}
Finally, the two expectation values for two PF $\{T_\alpha\left( \frac{t}{n_i} \right), \bar{T}_\alpha\left( \frac{t}{n_i} \right) \}$ are written as
\begin{align*}
\left\langle O(t)\right\rangle_{T_\alpha^{n_i}}&=\langle O(t)\rangle_U+\sum_{q=\alpha+1}^\infty t^q \sum_{l,l'}\frac{1}{n_i^{q-(l+l')}}\left(\left\langle U^\dagger (t)OC^{(q)}_{l,l'}\right\rangle+\textrm{h.c.}\right)\\&\phantom{=\;}+\sum_{q=\alpha+1}^\infty t^q\sum_{l,l',\bar{l},\bar{l}'}\frac{1}{n_i^{q-(l+l'+\bar{l}+\bar{l}')}}\sum_{q'=\alpha+1}^{q-(\alpha+1)}\left\langle (C^{(q')}_{l,l'})^\dagger O C^{(q-q')}_{\bar{l},\bar{l}'}\right\rangle\\
\left\langle O(t)\right\rangle_{\bar{T}_\alpha^{n_i}}&=\langle O(t)\rangle_U+\sum_{q=\alpha+1}^\infty t^q \sum_{l,l'}\frac{(-1)^{q-(l+l')}}{n_i^{q-(l+l')}}\left(\left\langle U^\dagger (t)OC^{(q)}_{l,l'}\right\rangle+\textrm{h.c.}\right)\\&\phantom{=\;}+\sum_{q=\alpha+1}^\infty t^q\sum_{l,l',\bar{l},\bar{l}'}\frac{(-1)^{q-(l+l'+\bar{l}+\bar{l}')}}{n_i^{q-(l+l'+\bar{l}+\bar{l}')}}\sum_{q'=\alpha+1}^{q-(\alpha+1)}\left\langle (C^{(q')}_{l,l'})^\dagger O C^{(q-q')}_{\bar{l},\bar{l}'}\right\rangle.
\end{align*}
Then, the average of two expectation values becomes
\begin{align*}
\frac{1}{2}\left(\left\langle O(t)\right\rangle_{T^{n_i}}+\left\langle O(t)\right\rangle_{\bar{T}^{n_i}}\right)&=\langle O(t)\rangle_U +\frac{1}{2}\sum_{q=\alpha+1}^\infty t^q \sum_{l,l'}\frac{1+(-1)^{q-(l+l')}}{n_i^{q-(l+l')}}\left(\left\langle U^\dagger(t) OC^{(q)}_{l,l'}\right\rangle+\textrm{h.c.}\right)\\&\phantom{=\;}+\frac{1}{2}\sum_{q=\alpha+1}^\infty t^q\sum_{l,l',\bar{l},\bar{l}'}\frac{1+(-1)^{q-(l+l'+\bar{l}+\bar{l}')}}{n_i^{q-(l+l'+\bar{l}+\bar{l}')}}\sum_{q'=\alpha+1}^{q-(\alpha+1)}\left\langle (C^{(q')}_{l,l'})^\dagger O C^{(q-q')}_{\bar{l},\bar{l}'}\right\rangle,
\end{align*}
which includes polynomials of $1/n_i^{\alpha+1},1/n_i^{\alpha+3},\cdots ,1/n_i^{q-1}$, for $\alpha\in\mathbb{Z}_{\rm odd}$, as an Trotter error for both $t^q$ and $t^{q+1}$ terms for any given odd $q$. Therefore, the coefficients $\{c_i\}$ satisfying Eq.~\eqref{eq:vander_dual}, and it completes our proof.

\end{document}